\documentclass[conference]{IEEEtran}
\IEEEoverridecommandlockouts
\usepackage{amsmath,amssymb,amsfonts}
\usepackage{algorithmic}
\usepackage{graphicx}
\usepackage{textcomp}
\usepackage[table,xcdraw]{xcolor}
\usepackage{xcolor}

\usepackage{hyperref}
\usepackage{biblatex}
\usepackage{color}
\usepackage{multirow}
\def\BibTeX{{\rm B\kern-.05em{\sc i\kern-.025em b}\kern-.08em
    T\kern-.1667em\lower.7ex\hbox{E}\kern-.125emX}}
\DeclareUnicodeCharacter{2061}{}

\usepackage{comment}
\usepackage{booktabs}
\usepackage{subcaption}
\usepackage{url}
\usepackage{bm}

\begin{document}

\title{How Does Imperfect Automatic Indexing Affect Semantic Search Performance?\\
}

\author{
\IEEEauthorblockN{Mengtian Guo, David Gotz, Yue Wang}
\IEEEauthorblockA{\textit{School of Information and Library Science} \\
\textit{University of North Carolina at Chapel Hill}\\
Chapel Hill, NC, USA \\
\{mtguo, gotz, wangyue\}@unc.edu}
}

\maketitle

\begin{abstract}
Documents in the health domain are often annotated with semantic concepts (i.e., terms) from controlled vocabularies. As the volume of these documents gets large, the annotation work is increasingly done by algorithms. Compared to humans, automatic indexing algorithms are imperfect and may assign wrong terms to documents, which affect subsequent search tasks where queries contain these terms.
In this work, we aim to understand the performance impact of using imperfectly  assigned terms in Boolean semantic searches. We used MeSH terms and biomedical literature search as a case study. We implemented multiple automatic indexing algorithms on real-world Boolean queries that consist of MeSH terms, and found that (1) probabilistic logic can handle inaccurately assigned terms better than traditional Boolean logic, (2) query-level performance is mostly limited by lowest-performing terms in a query, and (3) mixing a small amount of human indexing with automatic indexing can regain excellent query-level performance. These findings provide important implications for future work on automatic indexing.
\end{abstract}


\section{Introduction}
Controlled vocabulary is widely used in search engines to index unstructured content with semantic concepts. For instance, Medical Subject Headings (MeSH) are developed by the National Library of Medicine (NLM) for indexing and searching biomedical information in the PubMed search engine. Other medical search engines such as SemEHR~\cite{wu_semehr_2018} and Thalia~\cite{soto_thalia_2019} assign terms in the Unified Medical Language System (UMLS) to documents and use these terms as search facets. The radiology image search engine GoldMiner~\cite{kahn_goldminer_2007} assigns terms in SNOMED-CT and MeSH to image documents to facilitate image search. 

The rapid growth of biomedical information and clinical texts makes it infeasible to manually assign controlled vocabulary terms to each document. In 2022, more than 1.3 million articles were added to the PubMed index~\cite{medline_statistics}. In the same year, NLM announced that all PubMed articles have been indexed by automatic methods named the Medical Text Indexer (MTI)~\cite{nlm_medical_text_indexer}, with only a small subset reviewed and curated by human experts~\cite{automated_indexing_faqs}. 

However, the accuracy of state-of-the-art automated indexing methods still cannot match that of human indexing~\cite{mork201712}. In BioASQ 2022, a competition that focused on MeSH indexing, the best system achieved a micro-F1 score below 0.75~\cite{nentidis2022bioasq}, on par with or better than NLM's system in the same competition. Indeed, NLM is aware that automatic indexing have potential errors~\cite{automated_indexing_faqs}. Wrongly indexed articles may result in unsatisfactory search results. For example, in systematic literature reviews, researchers compose Boolean queries to precisely define the inclusion and exclusion criteria in a search query, and these queries usually contain MeSH terms to specify topics of interest. Automatic indexing algorithms may make mistakes in assigning MeSH terms to articles, causing relevant articles missed and non-relevant articles included in search results. 

Given the inevitable trend of adopting automatic indexers in semantic search engines, we are interested in answering the following question: \textbf{how will inaccurate automatic indexing results influence semantic search performance, where queries are Boolean combinations of index terms?} For search engine developers, it is important to systematically measure the influence of adopting automatic indexing techniques on search quality before deploying them in production. For search engine users, they generally expect terms to be precisely assigned to documents. As semantic search engines automate the indexing process, they should be informed to what extent this expectation still holds true.

\begin{figure}[t]
\centering
\includegraphics[width=0.8\columnwidth]{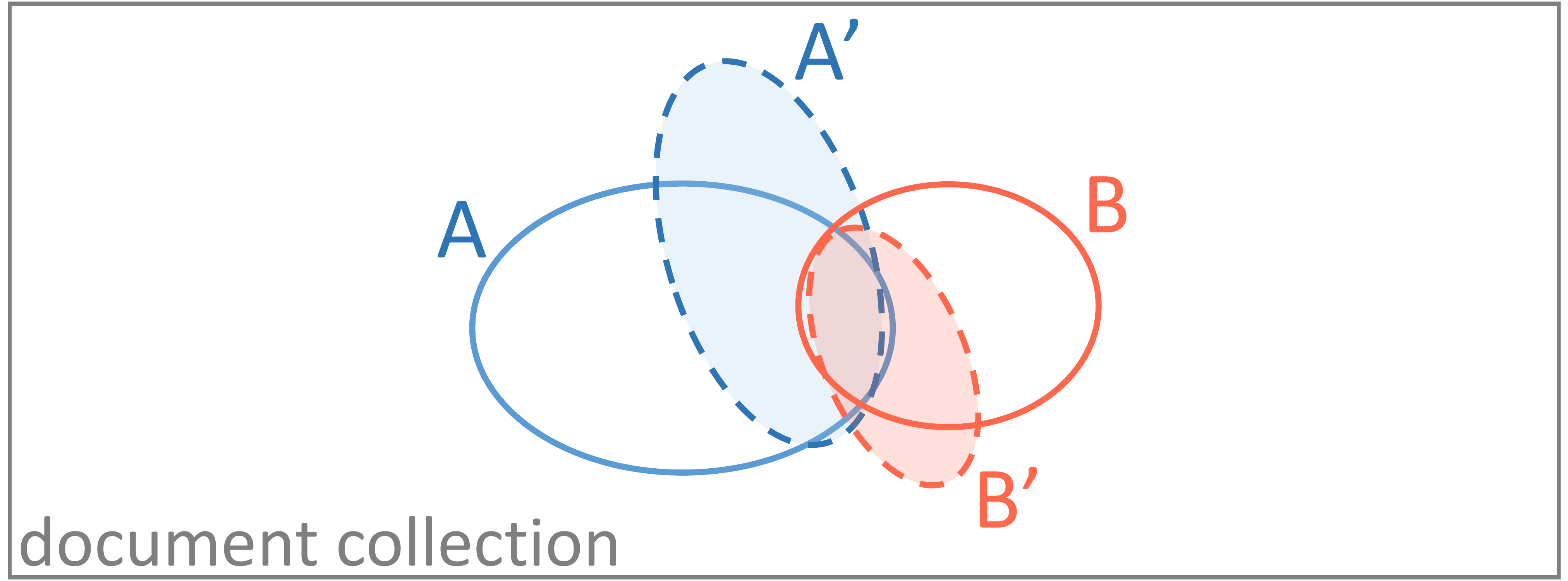}
\caption{A Venn diagram that illustrates the ground-truth article sets for two terms $A, B$ and the corresponding machine-classified article sets $A', B'$. The term-level classifiers are both inaccurate ($A \not= A', B \not= B'$). The query-level performance is good if the query is $A$ {\sc{and}} $B$ ($A \cap B \approx A' \cap B'$) but poor if the query is $A$ {\sc{or}} $B$  ($A \cup B \not= A' \cup B'$).}
\label{fig:venn}
\end{figure}

Answers to the above question may not be straightforward. To illustrate, consider the  example in Figure~\ref{fig:venn}. Here, the two term-level classifiers (automatic indexers) do not perform well individually. However, when used in a conjunctive Boolean query, the resulting article set matches the ground-truth set very well. Meanwhile, the resulting article set of a disjunctive query severely mismatches the ground truth. This example shows that inaccurate term-level indexers may or may not give poor query-level results. The influence depends not only on the accuracy of term-level classifiers, but also on how the terms are logically combined in queries. 

In this work, we take a data-driven approach to investigating the effects of automatic indexing on semantic search quality. Specifically, we study the case of MeSH indexing and biomedical literature search. Using a set of real-world search queries consisting of MeSH terms, we compared the search results using automatically assigned MeSH terms against search results using manually assigned MeSH terms. We considered two automatic indexing approaches: assigning MeSH terms as binary labels (presence or absence) or probabilistic labels (each term is associated with a predicted probability). Correspondingly, search queries are implemented as either Boolean logic or probabilistic logic. Our analyses on the impact of inaccurate indexing on search results reveal the following major findings:
\begin{itemize}
\item Assigning terms as probabilistic labels delivers better search performance than assigning terms as binary labels. Keeping the predicted probability of each term makes it possible to incorporate term-level uncertainty information in subsequent search and ranking processes.
\item Query-level search performance does \emph{not} linearly depend on term-level classification performance. Rather, we see a `bucket effect': query-level performance is limited by the \emph{worse-performing terms} in the query, which are often rare terms in the corpus.
\item Excellent query-level performance ($>95$\% F1) can be achieved by dividing the indexing work between humans and machines: to apply automatic indexing only on terms that are relatively easy for machines to predict ($>80$\% F1), and use manual indexing on remaining (rare) terms that are challenging for machines to predict.
\end{itemize}

Our work reveals several implications for future research on automatic indexing. It shows that index terms with lowest predictive performance are the bottlenecks of semantic search performance. It is therefore crucial for automatic indexing research to specifically focus on optimizing rare terms. At the same time, investing human reviewing efforts on these terms may yield a high return in search performance gain.

\section{Related Work}
To the best of our knowledge, the relationship between term-level indexing performance and query-level search performance has not been well-studied. Prior works often evaluate automatic indexing techniques at the term or concept level by treating the indexing problem as a multi-label classification or information extraction problem~\cite{kahn_automated_2009, neveol_evaluation_2005} . Few works have evaluated automatic indexing systems in the context of search tasks. An early work by Kim et al.~\cite{kim_automatic_2001} studied the impact of automatically assigned MeSH terms on search results against manually assigned MeSH terms. They measured the impact using a set of non-Boolean queries. However, Boolean queries are where MeSH terms are often used in practice~\cite{lefebvre2019searching}.

Specific efforts have been put into extracting biomedical concepts from literature, often referred as semantic tagging or semantic indexing. In a dedicated competition, BioASQ Task A \cite{tsatsaronis_overview_2015}, various methods have been proposed to index biomedical publications with MeSH terms automatically \cite{you_bertmesh_2021, xun_meshprobenet_2019}. This line of work usually evaluates semantic indexing systems as solving an extreme multi-label classification problem.

The problem of automatic MeSH indexing is broadly related to information extraction, an active research area that focuses on extracting valuable information from unstructured data. Information extraction services are appealing to biomedical knowledge management systems in that they can produce structured, unambiguous data from unstructured text, thus enabling more effective and efficient search \cite{jovanovic_automated_2014}. Therefore, it has been well-studied and applied to systems in various applications such as EHR systems \cite{wu_semehr_2018}, image retrieval \cite{kahn_automated_2009}, and biomedical literature search \cite{taboada_automated_2014}. Traditionally, machine-extracted terms are not directly exposed to users but rather internally used by search engines to understand search queries and improve search quality~\cite{reinanda_knowledge_nodate}. Recent systems such as PubMed start to allow end-users to directly express their search and filter criteria through machine-assigned index terms~\cite{automated_indexing_faqs}.

\section{Methods}
\subsection{Problem Formulation}

\textbf{Document indexing}.  Let the set of all MeSH terms be $H = \{h_1, \cdots, h_n\}$. The indexing process assigns terms from $H$ to a document $d$. The ground-truth/manual term assignment vector for $d$ is $s = (s_1, \cdots, s_n)$. $s_i=1$ indicates that $h_i$ was assigned to $d$, and $s_i=0$ indicates otherwise. 

Automatic indexing algorithms (multi-label classifiers or index term rankers) assign terms to $d$ as either binary labels or probabilistic labels. A binary indexing algorithm assigns a binary vector $b = (b_1, \cdots, b_n)$ to $d$. $b_i \in \{0,1\}$ indicates whether the algorithm predicts that $h_i$ belongs to $d$ or not. A probability indexing algorithm assigns a probability vector $p = (p_1, \cdots, p_n)$ to $d$. $p_i \in [0,1]$ indicates the predicted probability that $h_i$ belongs to $d$. 

\textbf{Boolean query using index terms}. A Boolean query $q$ is a  combination of  MeSH terms to express a relevance criterion.
The Boolean combination is a function $f_q(s)$ that takes the term assignment vector $s$ and outputs a binary relevance value. If $d$'s ground-truth term assignment vector is $s$, then $f_q(s) \in \{0,1\}$ is the ground-truth for whether $d$ is relevant to $q$.  Table \ref{tab:example_simplified} shows an example of a search topic and the corresponding formulation. 

When manual indexing is replaced by automatic indexing,  we can estimate the relevance of $d$ to a query $q$ based on the terms automatically assigned to $d$. For a binary indexing algorithm that gives a binary vector $b$, $f_q(b) \in \{0,1\}$ is the estimated binary relevance for whether $d$ is relevant to $q$.
For a probability indexing algorithm that gives a probability vector $p$, we can use the probabilistic version of the Boolean function to estimate the probability that $d$ is relevant to $q$. We call this probabilistic logic function $f'_q$. In this paper, we implement the probabilistic logic using product $t$-norm~\cite{hajek_metamathematics_1998}. The last row of Table \ref{tab:example_simplified} shows such an example.

\begin{table*}[htbp!]
\centering
\caption{An example that illustrates a search topic represented as a Boolean logic or probabilistic logic over MeSH terms. `$\oplus$' is the logical {\sc{or}} and `$\odot$' is the logical {\sc{and}}. }
\small
\begin{tabular}{l|l}
\toprule
Search topic                         & ``COVID-19 transmission''                                                                      \\ \midrule
Boolean search query                 & $q$ = (``COVID-19'' {\sc{or}} ``SARS-CoV-2'' {\sc{or}} ``Coronavirus'') {\sc{and}} ``Disease Transmission, Infectious''      \\ \midrule
MeSH terms in $q$                 & \begin{tabular}{@{}p{0.6\textwidth}@{}}$h_1$ = ``COVID-19'' (D000086382), $h_2$ = ``SARS-CoV-2'' (D000086402), \\ $h_3$ = ``Coronavirus'' (D017934), $h_4$ = ``Disease Transmission, Infectious'' (D018562)\end{tabular} \\ \midrule
Boolean logic function $f_q$              & \begin{tabular}{@{}p{0.6\textwidth}@{}}$f_q (b) = (b_1\ \oplus b_2\ \oplus b_3)\ \odot b_4$  \end{tabular}                                                       \\ \midrule
Probabilistic logic function $f'_q$ & \begin{tabular}{@{}p{0.6\textwidth}@{}} 
    $f'_q (p) = [(p_1 + p_2 - p_1 \cdot p_2) + p_3 - (p_1 + p_2 - p_1 \cdot p_2)\cdot p_3]\cdot p_4 $
\end{tabular}                                                         \\ \bottomrule
\end{tabular}
\label{tab:example_simplified}
\end{table*}

\subsection{Indexing Algorithms}
\label{sec:index_algorithms}
We implemented automatic indexing as a multi-label classification task, where the document content (title and abstract) is the input and a set of associated MeSH terms is the output. We trained two types of MeSH term classifiers: logistic regression (LR) and Bidirectional Encoder
Representations from Transformers (BERT). The simpler LR classifiers used bag-of-words features with TF-IDF weights. We trained one binary LR classifier for each MeSH terms.  The more complex BERT classifier was obtained by fine-tuning BioBERT~\cite{lee_biobert_2019} with as many sigmoid units as the number of MeSH terms. Both classifiers produce a vector of predicted probabilities for each term. We obtain a vector of binary values for each term by thresholding the predicted probabilities at $0.5$. 

The two classifiers give us four algorithms for query-level document search. If their outputs are probabilities for each term (before thresholding), we use probabilistic logic to combine these probabilities to obtain a query-level relevance score in $[0,1]$. We call the two algorithms \textbf{Prob}$_\text{LR}$ and \textbf{Prob}$_\text{BERT}$. 
If their outputs are binary values for each term (after thresholding), we use Boolean logic to combine these binary values to obtain a query-level binary relevance score in $\{0,1\}$. We call the two algorithms \textbf{Binary}$_\text{LR}$ and \textbf{Binary}$_\text{BERT}$.

As a comparison, we also trained classification models that take a document $d$ as input and \emph{directly} predict the query-level ground-truth label $f_q(s)$ without predicting the term-level label vector $s$. In this method, the model can be optimized in an end-to-end fashion to directly improve the query-level performance and therefore may outperform the above pipeline models~\cite{wang_deep_2019}. Therefore, we include this method as a comparison. For each query, we train both LR and BERT models to predict the probability of a document being relevant to a query. We call these algorithms \textbf{Query}$_\text{LR}$ and \textbf{Query}$_\text{BERT}$.  

\textbf{Prob}$_\text{LR}$, \textbf{Prob}$_\text{BERT}$, \textbf{Query}$_\text{LR}$, and \textbf{Query}$_\text{BERT}$ output continuous scores in $[0,1]$. Therefore they can either classify or rank documents for a given query. \textbf{Binary}$_\text{LR}$ and \textbf{Binary}$_\text{BERT}$ output binary values and can only generate classifications but not rankings of documents for a given query.

We did not use NLM's MTI algorithm or other algorithms from BioASQ competitions because our primary goal is to study the effect of using imperfectly assigned terms in Boolean semantic search. Those  algorithms are not qualitatively different from models we trained in our study in terms of their imperfect indexing performance. So instead we trained our own models to facilitate experimentation and comparison.

\subsection{Dataset Construction}
\textbf{Document indexing corpus}. We trained and evaluated the above LR and BERT models using a large and representative set of biomedical literature with MeSH term annotations from BioASQ 2021 Task A \cite{nentidis_overview_2021}. It contains 15,559,157 PubMed articles with titles, abstracts, and expert-annotated MeSH terms, among other metadata. The corpus covers 29,369 unique MeSH terms. Each article contains 12.68 MeSH terms on average. 

\textbf{Boolean query set}. To evaluate the influence of automatic indexing on search result quality, we used a set of Boolean queries constructed for systematic literature reviews. The queries are from three sources: (1) CLEF-TAR 2019~\cite{kanoulas2019clef},  (2) Systematic Review Update Dataset~\cite{alharbi_dataset_2019}, (3) PubMed COVID-19 article filters~\cite{noauthor_help_nodate}. Since the goal here is to obtain Boolean queries consisting of MeSH terms, not to perform the same systematic review tasks \emph{per se}, we processed the queries as follows. First, we eliminated any non-MeSH terms in these queries so that each query is a logical composition of MeSH terms only. Second, if a term is preceded by the exp (explosion) operator, we only included the term itself without expanding it to its hyponyms in the MeSH hierarchy. This is because such expansion only refines the scope of a term, not the semantic complexity of the query. Finally, we eliminated queries with less than 50 relevant documents in the entire corpus such that reasonably good machine learning models can be trained. The final query set includes 27 queries that contain 183 distinct MeSH terms. Each query combines a set of MeSH terms by logic operators ({\sc{and}}, {\sc{or}}, {\sc{not}}). All queries and references to their origins can be found in the \href{https://bit.ly/3mKmKIp}{online appendix}.\footnote{\url{https://bit.ly/3mKmKIp}}

\subsection{Implementation Details}
We split the dataset into training set (89\%; 13,847,650 articles), validation set (9\%; 1,400,324 articles), and test set (10\%; 1,555,915 articles). We randomly sampled a training subset with 500K documents from the training set. Regardless of which indexing approach is followed, imbalanced class distribution is a problem that the classifier training process must deal with. This is because only a small minority of the overall document corpus matches any given MeSH term or query. This imbalanced class distribution generally presents a challenge for machine learning algorithms. To mitigate this challenge, we up-sampled for rare MeSH terms and sparse queries by adding extra documents that are assigned with rare terms or are relevant to sparse queries to the training subset. We constructed such a training subset with 514,686 articles for the experiments. For a fair comparison, all classifiers were trained using the same set of articles. We found that using a larger training subset is not improving the MeSH terms predictors significantly.

Note that in this subset the positive examples of any class are still the minority. When training a machine learning classifier on a dataset with an imbalanced distribution of positive vs. negative examples, the classifier tends to be biased towards the majority (negative) class and reluctant to predict the minority (positive) class. To mitigate this problem, we used special loss functions such as the cost-sensitive loss \cite{he_learning_2009} and the focal loss \cite{lin_focal_2017} to counter the effects of data imbalance. See Appendix for detailed data sampling procedure~\ref{sec:account_for_imbalance}, hyperparameter settings~\ref{sec:hyperparameter}, and term-level classification performance (\ref{sec:clf_performance}).

\subsection{Query-level Evaluation Metrics}
Predicting the relevance of an article with respect to a query can be viewed both as a classification task (a binary notion of relevance) and a ranking task (a relative notion of relevance). To evaluate classification performance, we used precision, recall, and F1-score averaged over all 27 queries. To evaluate ranking performance, we used mean average precision (MAP), precision@10 (P@10), precision@50 (P@50), and recall@1000 (R@1000) averaged over all 27 queries. For the \textbf{Prob} and \textbf{Query} approaches, we converted continuous relevance scores into binary classification decisions by finding the threshold that maximizes the F1-score on the validation set, an approach proposed by You et al~\cite{you_bertmesh_2021}. We do not evaluate the ranking performance of the \textbf{Binary} approaches as they only generate binary relevance labels.

\section{Results and Analyses}

In this section, we conduct in-depth analyses of the performance of various automatic indexing and ranking algorithms described in Section \ref{sec:index_algorithms}. Besides comparing  different algorithms, we also observe a large performance variation across queries (\ref{sec:quant_perf_eval}). This motivates us to further analyze the correlation between term-level performance and query-level performance (\ref{sec:perf_corr_analysis}), and look into concrete cases (\ref{sec:case_study}). Finally, we analyze a mixed indexing scenario where algorithms only assign  terms that can be accurately predicted and humans assign terms that are challenging for algorithms to predict (\ref{sec:mixed_indexing}).

\subsection{Quantitative Performance Evaluation}
\label{sec:quant_perf_eval}

\begin{table*}[htbp!]
\centering
\caption{Query-level average performance of automatic indexing and ranking algorithms. All metrics are the higher the better. Algorithms using the same base classifier (LR or BERT) are grouped into adjacent rows. Within either group of rows in the same column: results labeled with `$^*$' are significantly better than results without `$^*$'; if two results are labeled with `$^*$', they are not significantly different. Across all rows in the same column: if a result is shown in \textbf{boldface}, it is significantly better than any other results in that column; otherwise all results labeled with `$^*$'  in that column do not have significant difference. (Randomization test, significance level $\alpha = 0.05$)}
\small
\begin{tabular}{llllllll}
\toprule
Method             & P     & R     & F1    & MAP   & P@10  & P@50  & R@1000 \\ \midrule
\textbf{Binary}$_\text{LR}$   & 0.310* & 0.625* & 0.390* & -     & -     & -     & -      \\
\textbf{Prob}$_\text{LR}$     & 0.325* & 0.584 & 0.395* & 0.415* & 0.600* & 0.459* & 0.791*  \\
\textbf{Query}$_\text{LR}$    & 0.241 & 0.543 & 0.305 & 0.314 & 0.507 & 0.401 & 0.745  \\ \midrule
\textbf{Binary}$_\text{BERT}$ & 0.358 & 0.553* & 0.391 &-       &-       &-       &-        \\
\textbf{Prob}$_\text{BERT}$   & \textbf{0.427*} & 0.554* & \textbf{0.457*} & 0.452* & 0.611 & 0.481 & 0.810*  \\
\textbf{Query}$_\text{BERT}$  & 0.389 & 0.434 & 0.389 & 0.397 & 0.530 & 0.471 & 0.711  \\ \bottomrule
\end{tabular}
\label{tab:main}
\end{table*}

We report query-level performance of the two semantic indexing methods (\textbf{Binary}$_\text{LR/BERT}$; \textbf{Prob}$_\text{LR/BERT}$) and the comparison method (\textbf{Query}$_\text{LR/BERT}$) in Table \ref{tab:main}.  To facilitate comparison between methods that used the same base classifiers, we put LR results and BERT results as adjacent rows. All metrics are averaged over 27 queries. We compared the metrics against each other using two-sided randomization test \cite{basu_randomization_1980} with Type-I error level at $\alpha=0.05$.
We seek to understand how much the search performance drops when using automatic indexing methods. We also seek to compare the performance difference across different indexing methods, different base models, and when evaluating on different search queries.

Since we treat the set of documents retrieved using manual indexing as the ground truth, the precision, recall, and F1 score of manual indexing should all be $1.0$. Overall, none of the experimented indexing methods perform closely to manual indexing. The best F1 score is achieved by \textbf{Prob}$_\text{BERT}$, with the best precision and a decent recall. \textbf{Binary}$_\text{LR}$ achieves the highest recall (although not significantly better than other recall results) at the cost of a low precision. \textbf{Prob}$_\text{BERT}$ gives the best ranking performance in all four metrics (although none of the metrics are significantly higher than those of \textbf{Prob}$_\text{LR}$).

Comparing \textbf{Binary} and \textbf{Prob} methods, \textbf{Prob} methods perform at least as well as \textbf{Binary} methods in most of the metrics except for recall when using LR as the base classifier. \textbf{Prob} methods show its advantage in precision when using BERT as the base classifier. Treating F1 as the measure for overall model performance, \textbf{Prob} performs similarly to \textbf{Binary} when using LR and significantly better than \textbf{Binary} when using BERT. \textbf{Prob} methods have the additional advantage of generating document rankings, making them more favorable than \textbf{Binary} methods for automatic indexing.

Comparing LR and BERT, there are no significant differences between \textbf{Binary}$_\text{LR}$ and \textbf{Binary}$_\text{BERT}$. When implemented as \textbf{Prob} indexers, \textbf{Prob}$_\text{BERT}$ gives significantly better query-level precision and F1. The recall and ranking metrics are similar between \textbf{Prob}$_\text{BERT}$ and \textbf{Prob}$_\text{LR}$.

The term-level indexing methods (\textbf{Binary}$_\text{LR/BERT}$; \textbf{Prob}$_\text{LR/BERT}$) perform significantly better than query-level classifiers (\textbf{Query}$_\text{LR/BERT}$) in most of the metrics. This means when the decision logic is pre-specified, end-to-end training is not necessarily better than component-wise training.
Therefore, term-level indexing is a good strategy for semantic search not only for its scalability and transparency but also for its performance. 

\begin{figure}[!h]
\centering
    \begin{subfigure}[t]{0.24\textwidth}
    \centering
    \includegraphics[width=\textwidth]{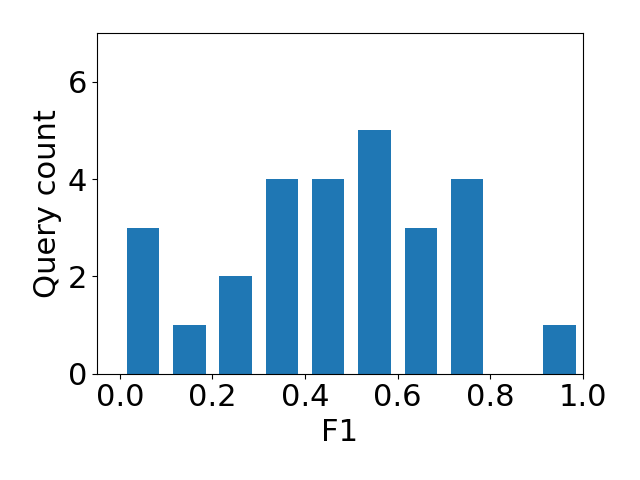}
    \caption{\textbf{Prob}$_\text{BERT}$} \label{fig:f1_hist_prob}
\end{subfigure}\hfill
\begin{subfigure}[t]{0.24\textwidth}
    \centering
    \includegraphics[width=\textwidth]{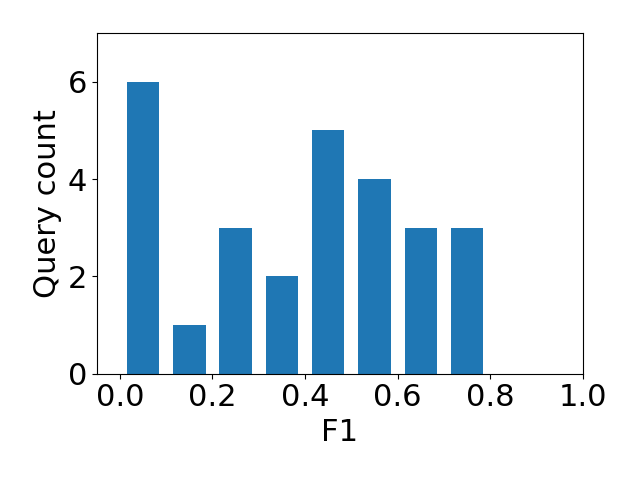}
    \caption{\textbf{Binary}$_\text{BERT}$} \label{fig:f1_hist_bi}
\end{subfigure}
\caption{F1 score distribution in two methods.}
\label{fig:F1}
\end{figure}

Looking at the performance distribution in Figure \ref{fig:F1}, the drop in performance varies among queries. Some queries are less influenced by automatic indexing and reach an F1 of 0.9016. Some queries have a big drop in performance and the F1 is 0.0 when using automatic indexing. This motivates the analysis of term-level factors that influence query performance and the case studies below.

\subsection{Correlation Between Term-Level Performance and Query-Level Performance}
\label{sec:perf_corr_analysis}

As observed in the query performance distribution in Figure \ref{fig:F1}, the impact of automatic indexing on queries varies. There are two possible sources of difference. First, term-level performances of the MeSH terms used in the queries are different, which result in differences in query-level performance. Second, the queries have different semantic structures and complexity (e.g. they use a different number of MeSH terms, and the MeSH terms are combined in different ways). To understand what factors result in the difference, we analyzed the correlation between the search result quality on a query and (1) aggregated term-level performance of MeSH terms used in the query and 2) query characteristics describing its complexity. We used \textbf{Prob}$_\text{BERT}$ in our analysis as it gives the overall best performance among all methods considered.

\subsubsection{Correlation between term-level performance and query-level performance}
We used F1 metric to represent the performance of a MeSH term indexer. We calculated the Pearson correlation between the classification performance on a query and the aggregated performance of MeSH terms used in that query (i.e. worst/average/best term-level F1, frequency of the most frequent MeSH term,  frequency of the least frequent MeSH term, and number of MeSH terms in that query). The results are shown in Table \ref{tab:analysis}. 

\begin{table*}[h]
\caption{Pearson correlation  between query-level performance (precision and recall) and different term-level characteristics. `$^*$' indicates a significant correlation ($p< 0.05$) and `$^\dagger$' indicates a moderately significant correlation ($0.05 \le p < 0.1$).}
\centering
\small
\begin{tabular}{lllllllll}
\toprule
                     & Worst F1                 & Avg F1                   & Best  F1 & Highest freq. & Lowest freq.   & \# of terms                 & \# of \textsc{And}'s                  & \# of \textsc{Or}'s                                  \\\midrule
Query-level precision                    & 0.290                         & 0.370$^\dagger$ & 0.321        & 0.196             & 0.140     & -0.226  & -0.137  & -0.222                                              \\
Query-level recall                    & 0.387$^*$ & 0.366$^\dagger$ & -0.033       & -0.139            & 0.403$^*$  &  -0.378$^\dagger$  & -0.144  & -0.377$^\dagger$
 \\\bottomrule
\end{tabular}
\label{tab:analysis}
\end{table*}

For the correlation between query performance and MeSH term performance, the recall on a query is significantly higher if the worst F1 among all the MeSH terms is high. Besides, the recall and precision on a query are higher if the average F1 of all the MeSH terms is higher (moderately significant). There is no significant correlation between the best F1 among all the MeSH terms with the precision and recall on a query. This indicates that the recall on a query is restricted by the worst-performing MeSH term indexers of MeSH terms in the query. To improve query retrieval performance, we should focus on  the lowest-performing MeSH terms and average MeSH term performance. Having one well-performing MeSH term in a query is not enough to give a good query-level search performance.

For the correlation between query performance and MeSH term frequency, we observe that the recall on a query is positively correlated with the frequency of the least frequent MeSH term. Therefore, if the query contains a rare MeSH term, it is more likely that the query will have a low recall. We found that the F1 score of a MeSH term has a positive correlation with its frequency in the training set (Pearson correlation: 0.218, $p$-value: 0.003). Therefore, the F1 score of the rare MeSH term might be low due to the lack of training instances, which brings down the retrieval performance on the query. Therefore, automatic indexing systems should put more efforts into improving the performance on rare MeSH terms.

\subsubsection{Correlation between query complexity and query-level performance}

The recall on a query is negatively correlated with the number of MeSH terms used in a query. One possible explanation is that queries with more MeSH terms are more likely to contain low-performing MeSH terms. We found that there is a strong correlation between the number of MeSH terms and the worst MeSH-level F1 in a query (Pearson correlation: -0.547, $p$-value: 0.003). Therefore, the more MeSH terms a query has, the more vulnerable it is to poor MeSH indexers. 

There is no significant correlation between the query-level performance and the number of \textsc{and}'s in the query. We found that the number of MeSH terms is strongly correlated with the number of \textsc{or}'s in a query (Pearson correlation: 0.977, p-value: 0.000), which indicates that when a query involves more MeSH terms, those MeSH terms are mostly connected by \textsc{or}. This in turn explains why we see a negative correlation between the query-level performance and the number of \textsc{or}'s in the query.

\subsection{Case Studies}
\label{sec:case_study}

To understand how the drop in search performance influences user's search, we provide three case studies below using the \textbf{Prob}$_\text{BERT}$ method. We selected three queries on which \textbf{Prob}$_\text{BERT}$ gives various performance: one query with $F1 = 0.7513$, one query with $F1 = 0.4583$, and one with $F1 = 0.0$. Imagine a clinician issues the three search queries in the testing set, which contains 1,555,915 articles. We compared the subset of articles retrieved using \textbf{Prob}$_\text{BERT}$ and using the ground-truth MeSH labels provided by humans. The set of articles retrieved using human-labeled MeSH terms are treated as relevant articles and those not retrieved are treated as non-relevant.
\subsubsection{Case 1}
For case study 1, we looked into the query related to clavicle fractures. The detailed query is shown below. On the search query, \textbf{Prob}$_\text{BERT}$ reaches a precision of 0.7184 and a recall of 0.7872. If the clinician issues the search query in the testing set, \textbf{Prob}$_\text{BERT}$ would retrieve 103 articles missing 20 relevant ones (e.g. PMID 32295588, PMID 28159682, PMID 27765500). For instance, PMID 28159682 (\textit{``Development of an injectable pseudo-bone thermo-gel for application in small bone fractures''}) describes a study of a drug developed to heal small bone fractures such as clavicle fractures. Among the 103 retrieved articles, 29 are non-relevant (e.g. PMID 30948397, PMID 31677623, PMID 31415405). For instance, PMID 30948397 (\textit{``Pucker sign in an adult distal radial fracture''}) mentioned that skin puckering is associated with clavicular fractures. However, the article provides a case of radial fracture instead of clavicular fractures. Both PMID 31677623 and PMID 31415405 were labeled with \textit{Clavicle}, and they were also assigned the \textit{Fracture Fixation, Internal} instead of \textit{Fracture Fixation}. Although they are not relevant to the given search query, they might still be relevant to the searcher's interests due to the relatedness of these two terms.
\begin{verbatim}
    1. Clavicle/
    2. Fractures, Bone/ or Fracture 
     Fixation/ or Fracture Healing/
    3. 1 and 2
    4. Animals/
    5. Humans/
    6. 4 not 5 
    7. 3 not 6
\end{verbatim}

\subsubsection{Case 2}
For case study 2, we looked into the query related to infant sepsis. On the search query, \textbf{Prob}$_\text{BERT}$ reaches a precision of 0.3738 and a recall of 0.5923. If the clinician issues the search query in the testing set, \textbf{Prob}$_\text{BERT}$ would
retrieve 618 articles missing 159 relevant papers (e.g. PMID 375620, PMID 918581, PMID 879115). Among the 618 retrieved articles, 387 articles are non-relevant to the search query (e.g. PMID 6798907, PMID 6787924, PMID 6783064).
\begin{verbatim}
    1. Sepsis/ and Infant, Newborn/
\end{verbatim}

\subsubsection{Case 3}
For case study 3, the search query is related to nutrition therapy of renal insufficiency. On this search query, \textbf{Prob}$_\text{BERT}$ has \emph{zero} precision and recall. Therefore, using automatic indexing, none of the 14 relevant articles (e.g. PMID 23123667, PMID 22177826, PMID 20708074) could be retrieved. \textbf{Prob}$_\text{BERT}$ retrieves 68 false positive papers (e.g. PMID 87126, PMID 113851, PMID 104103). These three papers are all labeled with ``Amino Acids'' and other MeSH terms potentially related to nutrition. However, none of them are labeled with ``Renal Insufficiency'' 
(the lowest-performing term in this query; F1 = 0.0057).

\begin{verbatim}
    1. Renal Insufficiency/
    2. Nutrition Therapy/
    3. Nutritional Requirements/
    4. Energy Intake/
    5. Infusions, Parenteral/
    6. Gastrostomy/
    7. Dietary Proteins/
    8. Amino Acids/
    9. Glucose/
    10. or/2-9
    11. 1 and 10
\end{verbatim}

\subsection{Dividing Indexing Work Between Humans and Machines}
\label{sec:mixed_indexing}

Our analyses above shows that the underperforming terms in automatic indexing are the major bottleneck of query-level search performance. Their low prevalence in the training set makes it intrinsically challenging for machine learning models to predict. On the other hand, manual indexing cannot  scale up to large document collections even though it ensures high search performance.  In this section, we seek to find the balance between full automatic indexing and full manual indexing. We imagine a mixed indexing scenario where we  use an algorithm to only assign terms that can be accurately predicted, and still rely on humans to assign terms on which an algorithm underperforms. 

\begin{figure}[t]
\centering
\includegraphics[width=.9\columnwidth]{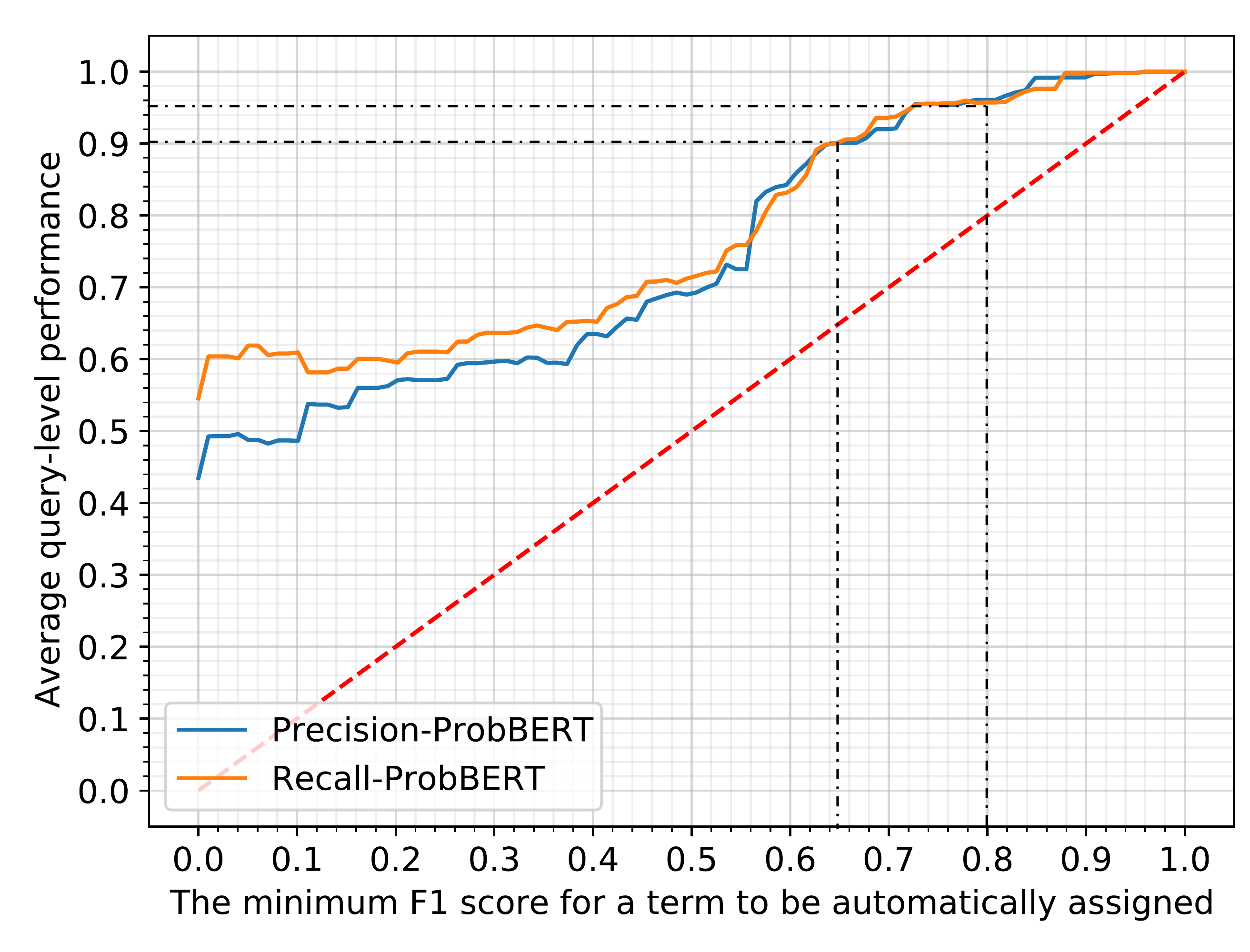}
\caption{Average query-level performance when replacing inaccurate automatic indexing results with ground-truth/manual indexing results. A point $(x,y)$ on a curve shows the average query-level performance is $y$ when  terms with F1 score $\ge x$ are automatically assigned and terms with F1 score $<x$ are manually assigned. The two dashed lines show that the average query-level precision/recall is 0.9 (0.95) when  terms with F1 score $\ge$ 0.65 (0.8) are automatically assigned and the rest terms are manually assigned, respectively.
}
\label{fig:replace}
\end{figure}

To simulate this mixed indexing scenario, we replace machine-assigned terms with human-assigned ground-truth terms if \textbf{Prob}$_\text{BERT}$ does not achieve a term-level F1 score higher than a threshold $t$. After the replacement, we re-evaluate the average query-level performance. When $t = 0$, no machine-assigned terms are replaced, and the query-level F1 equals the fully automatic result reported in Table~\ref{tab:main} (P = 0.427, R = 0.554). On the other hand, when $t = 1$, all terms are ground-truth assignments and the query-level performance is perfect (P = R = 1.0). As we increase $t$ from 0 to 1, a progressively larger underperforming subset of the 183 MeSH terms are replaced with ground-truth assignments, and we expect to see a monotonically increasing trend in query-level performance. The trend is shown in Figure~\ref{fig:replace}.

Figure~\ref{fig:replace} shows that in order to achieve above 0.9 (0.95) query-level precision/recall, one can use automatic indexing to assign terms that have above 0.65 (0.8) F1, respectively. The underperforming terms that have below 0.65 (0.8) F1 are manually assigned, which collectively account for 15.87\% (21.33\%) of MeSH term instances, respectively. An explanation for these relatively small fractions is that these underperforming terms are also \emph{rare} terms. 
This implies that an indexing algorithm does not need to be perfect to be useful in delivering high semantic search performance if automatic indexing methods are in charge of (1) assigning terms that have high predictive performance directly to documents, and (2) recommending terms that are underperforming but potentially relevant to assist humans in  making the indexing decision.

This analysis can be useful in two way. First, for technical research on MeSH term indexing algorithms (e.g., the BioASQ large-scale indexing challenge), it answers ``how good is good enough'' if the indexing workload is divided between algorithms and human experts. From a performance evaluation standpoint, the analysis highlights the importance of improving the predictive accuracy for those difficult terms, instead of placing equal importance on all terms.
Second, for semantic search service providers (e.g., PubMed), this analysis shows a methodology for determining which index terms can be automatically assigned without severely impacting index term-based search performance, and for quantifying the tradeoff between the cost of human review/curation vs. query-level search performance.

\subsection{Limitations}
A main limitation of the above analyses is that our experiment only used a small query set. The ideal query set is the log of search queries containing MeSH terms submitted to a real search engine such as the PubMed. We could not find such publicly available query logs, and therefore constructed a small query set. We believe the analysis methodology presented here is useful when such query logs are available. 

Another limitation is that in literature search, MeSH terms  are often used together with free-text terms. Our study considered Boolean searches that only contain MeSH terms without free-text terms, which are less used in literature search. However,  Boolean combinations of clinical terms are widely used to define complex high-level concepts in computational phenotyping~\cite{conway2011analyzing}and value sets ~\cite{willett2018snomed}. 

\section{Conclusion and Future Work}
In this work, we sought to understand the performance impact of using automatically assigned terms in semantic Boolean retrieval. We found that  probabilistic logic can better handle the uncertainty of predicted terms in document retrieval; query-level performance strongly correlates with underperforming terms in a query; and mixing a small amount of manual indexing into automatic indexing achieves excellent query-level performance. Although we used the specific case of MeSH terms and literature search,  the findings have implications for future work on automatic indexing in general.

\textbf{Incorporating  uncertainty of predicted terms in semantic Boolean search}: Automatic indexing algorithms often assign index terms with a degree of uncertainty, expressed through classifier-predicted probabilities or ranker-generated scores (which can be converted to predicted probabilities~\cite{niculescu2005predicting}).  Terms are then assigned to documents as binary labels by taking terms with a score above a threshold or taking the top $k$ terms. This procedure discards the inherent uncertainty in machine predictions. When these inaccurately assigned terms are used in Boolean searches, they are treated as if they were accurately assigned by humans, leading to subsequent retrieval errors. Our comparison of binary vs. probability indexing algorithms shows that it is beneficial to incorporate the uncertainty of predicted terms in subsequent Boolean searches by replacing Boolean logic with probabilistic logic. This naturally gives a ranked list of documents by following the probability ranking principle. 

\textbf{Emphasizing low-performance terms in automatic indexing algorithm training and evaluation}: Our analysis shows that low-performance terms are ``the shortest planks of a bucket'' -- they are the limiting factors of  query-level performance. Therefore, future research on automatic indexing algorithm training and evaluation should pay special  attention to the subset of terms that are rare and/or still suffer from relatively low predictive performance.  An emerging line of work on long-tail classes has been exploring similar challenges~\cite{xiao2021does}. For example, a high recall of the correct term among the top-$k$ can be a useful metric if the goal is to ask human indexers to review the list of potentially relevant candidate terms.

\textbf{Dividing indexing work between machines and humans}:
We advocate using manual indexing on terms where automatic indexing falls short.
This strategy was employed by NLM's MeSH term indexing before 2022~\cite{mork201712}. However, in NLM's most recent document, it is unclear how human review/curation efforts are combined with automatic term assignments in the `MTI-Auto' system~\cite{automated_indexing_faqs}. 
Our work shows that complementing automatic indexing with manual indexing can prevent low-performing terms from impacting query-level performance. Since low-performing terms are also rare, the associated manual indexing workload is relatively small compared to the entire indexing workload measured by the number of term assignment instances. In the long run, an increasing amount of labels for rare terms may become available. This will hopefully improve automatic indexing performance on those terms and the amount of human indexing work will gradually decrease.

\printbibliography

\appendix
\subsection{Accounting for Imbalanced Data Distribution}
\label{sec:account_for_imbalance}
\subsubsection{Upsampling Rare Concepts}
MeSH classifiers in the composable approach and query classifiers in the monolithic approach face the problem of imbalanced class distribution since positive examples are always orders of magnitude fewer than negative examples in the training set. The imbalanced class distribution generally presents a challenge for machine learning algorithms. To train high-quality classifiers, we used the following sampling method to incorporate more instances from rare classes. 

Suppose we have in total $M$ possible labels and $N$ documents in a sub-sample. If all labels were mutually exclusive, then each label should account for $N/M$ documents, called the ``fair share'' of a label. We define a class $c$ to be rare if, $N_c$, the number of documents associated with it, is smaller than 10\% of the ``fair share'' (i.e., $N_c < 0.1 \times \frac{N}{M}$). We then up-sample the rare classes to reach the 10\% ``fair share.'' In this study, $M$ is the total number of MeSH terms and queries (i.e., 183 + 27 = 210), and $N$ is the size of two training subsets (i.e., 500,000 and 1,000,000). After the sampling process, the sizes of the two training datasets are 514,686 (1.03 $\times$ 500,000) and 1,025,855 (1.03 $\times$ 1,000,000). We denote the two datasets as ``subset-500K'' and ``subset-1M''. We found that models trained on subset-1M are not significantly better than models trained on subset-500K. Therefore, we used models trained on subset-500K in the main text.

\subsubsection{Loss Functions for Imbalanced Class Distribution}
One challenge of training classifiers for MeSH and query prediction is the imbalance between positive and negative instances in the training dataset. The ratio between the positive and negative instances is from 1:0.5 to 1:11968 for MeSH terms and from 1:254 to 1:10090 for queries. We applied a weighting factor for the positive and negative class in the loss function to account for this. 

For Logistic Regression, we used the formula:
\begin{align}
    CE(p_t) = -\alpha_t  \log⁡(p_t) \\
    \alpha_t = \frac{n_\text{samples}}{n_\text{classes} \times n_\text{samples in class $t$}}
\end{align}
where $CE$ represents cross entropy loss; $p_t$ is the estimated probability of the ground-truth class ($p_0=1-p_1$); $\alpha_t$ is the weighting factor calculated through inverse class probability.

For BERT, we applied focal loss~\cite{lin_focal_2017} as it added stability to the training process. The following formula describes focal loss:
\begin{align}
    FL(p_t) = -\alpha_t (1-p_t )^\gamma \log⁡(p_t)
\end{align}
where $FL$ represents focal loss; $\gamma$ and $\alpha_t$ are hyperparameters set empirically, and we used the default values ($\gamma=2, \alpha_t=0.25$), which were found to be most effective in the paper that introduced focal loss.

\subsection{Model Hyper-parameter Settings}
\label{sec:hyperparameter}

\textbf{LR}. Logistic Regression models were trained through scikit-learn where we used L2 regularization, LBFGS optimizer, and a max iteration of 1,000. We set ``class\_weight'' to be ``balanced'' to add the weighting factor described in the previous section. All the other hyper-parameters are set to default.

\textbf{BERT}. We implemented BioBERT using Huggingface Transformers. We used grid-search for hyper-parameter tuning: learning rate \{5e-6, 1e-5, 2e-5, 3e-5\}, number of epochs \{1, 2, 3\}. We used a batch size of 8. We found that a learning rate of 1e-5 and three epochs gave the best performing models.

\subsection{MeSH Term Classifier Performance}
\label{sec:clf_performance}

\begin{table}[h!]
\caption{Macro-averaged classification performance (averaged over 183 MeSH terms in the experimental query set) on subset-500K and subset-1M.}
\centering
\small
\begin{tabular}{@{}lllll@{}}
\toprule
Data                         & Model & Precision & Recall & F1    \\ \midrule
\multirow{2}{*}{subset-500K} & LR    & 0.393     & 0.503  & 0.422 \\
                             & BERT  & 0.489     & 0.460  & 0.445 \\
\multirow{2}{*}{subset-1M}   & LR    & 0.393     & 0.538  & 0.434 \\
                             & BERT  & 0.487     & 0.444  & 0.431 \\ \bottomrule
\end{tabular}
\end{table}

\end{document}